\documentclass[9pt,twocolumn,twoside]{opticajnl}
\journal{opticajournal} % use for journal or Optica Open submissions

\setboolean{shortarticle}{true}
\usepackage{lineno}
\title{Temporal Talbot Effect: From a Quasi-Linear Talbot Carpet to Soliton Crystals and Talbot Solitons}

\author[1,*]{Marina Zajnulina}
\author[2]{Michael B{\"o}hm}

\affil[1]{Multitel Innovation Centre, Rue Pierre et Marie Curie 2, Mons, 7000 Belgium}
\affil[2]{Wismar University of Applied Sciences, Philipp-M{\"u}ller-Str., Wismar, 23966 Germany}

\affil[*]{zajnulina@multitel.be}

\begin{abstract}
The temporal Talbot effect refers to the periodic self-imaging of pulse trains in optical fibers. The connection between the linear and nonlinear temporal Talbot effect is still not fully understood. To address this challenge, we use Soliton Radiation Beat Analysis and numerically investigate the evolution of a phase-modulated continuous-wave laser input in a passive single-mode fiber. We identify three input-power-dependent regimes and their Talbot carpets: the quasi-linear regime for low input powers, the intermediate one, and separated Talbot solitons for higher powers. We show that the intermediate regime hosts soliton crystals rather than rogue waves, as reported in the literature. The Talbot-solitons beating can be used for pulse repetition-rate multiplication in the nonlinear regime. We also show two types of solitons involved: some encoded in the whole frequency comb and the individual solitons carried only by particular comb lines.
\end{abstract}

\setboolean{displaycopyright}{false} % Do not include copyright or licensing information in submission.

\begin{document}

\maketitle

\textit{Introduction.}
The spatial Talbot effect is a self-imaging phenomenon discovered by Henry Fox Talbot in 1836. If a monochromatic wave illuminates a periodic diffraction grating, the exact image (primary Talbot image) of this object reoccurs at the so-called Talbot length $L_{T}.$ At $L_{T}/2,$ the secondary Talbot image reappears with doubled frequency (i.e. halved grating period) and phase-shifted by $\pi.$ Further fractional Talbot images appear at regular fractions of $L_{T}$ constituting the so-called Talbot carpet. This effect is diffraction-driven and, as such, linear \cite{Wen_2013}. In 1981, Jannson \& Jannson report the temporal self-imaging in optical fibers \cite{Jannson_1981}. Since then, this linear dispersion-driven counterpart of the spatial Talbot effect, the so-called temporal Talbot effect, is widely used for lossless repetition-rate multiplication of optical pulses in fibers. The principle behind is the following: if a train of identical pulses with repetition rate of $\Omega$ enters a dispersive medium, its optical power image reoccurs after the Talbot period $L_{T}=\frac{2\pi}{(2\pi \Omega)^{2} |\beta_{2}|}.$ The fractional Talbot effect describes the reoccurrence of the sub-images at lengths $L_{p/q}=\frac{p}{q}\frac{2\pi}{(2\pi \Omega)^{2} |\beta_{2}|}$ with $p$ and $q$ being mutually prime natural numbers. The pulse repetition rate of these sub-images increases to $q\Omega,$ whereas the energy of each pulse is divided by $q$ \cite{Jannson_1981}, \cite{Azana_2003}, \cite{Cortes_2019}, \cite{Maram_2015}. 

The nonlinear spatial Talbot effect was reported in 2010 as the second-harmonic self-imaging of a periodic intensity pattern at the output facet of a periodically polled nonlinear crystal \cite{Wen_2011}. Spatial Talbot solitons have been generated via nonlinear interference of beams in bulk Kerr media \cite{Cohen_2008}. In the mid of 2010s, the temporal nonlinear Talbot effect in optical fibers gets into the discussion. Thus, in Refs. \cite{Zhang_2015a}, \cite{Zhang_2015b}, \cite{Nikolic_2019}, the authors construct doubly-periodic Akhmediev breathers as an initial condition to numerically integrate the Nonlinear Schr{\"o}dinger equation (NLS) for light propagation in fibers. They argue that the resulting Talbot carpets consist of rogue waves and point out that, in contrast to the linear temporal Talbot effect, these carpets lack higher-fractional images, only primary and secondary images are available. 

Although these studies represent a good starting point, there are still many open questions and challenges to address concerning the nonlinear temporal Talbot effect. Thus, the observed spatial and temporal (deterministic) regularity of the reported Talbot carpets contradicts the definition of rogue waves as (probabilistic) rare extreme events (\cite{Zhang_2015a}, \cite{Zhang_2015b}, \cite{Nikolic_2019}). Therefore, the question is whether other types of solitonic waves rather than rogues waves (often described as Peregrine solitons \cite{Tikan_2017}) are involved in the formation of nonlinear Talbot carpets. Also, whereas the construction of doubly-periodic Akhmediev breathers as initial conditions is numerically a straightforward task, experimentally, it is challenging. Are there other, more experimentally accessible, possibilities to create initial conditions for nonlinear Talbot carpets? Most importantly, how does the nonlinear temporal Talbot effect result from the linear one? Here, we address these questions. To begin with, using a phase-modulated continuous-wave (CW) laser input with a corresponding frequency comb in its spectrum can replace the construction of doubly-periodic Akhmediev breathers as initial conditions both experimentally \cite{Wu_2022} and numerically as we show later. Using Soliton Radiation Beat Analysis (SRBA), we answer the question of the connection between the linear and nonlinear Talbot effect and the soliton types involved. 

\begin{figure*}[th]
\centering
\fbox{\includegraphics[width= 0.95\textwidth]{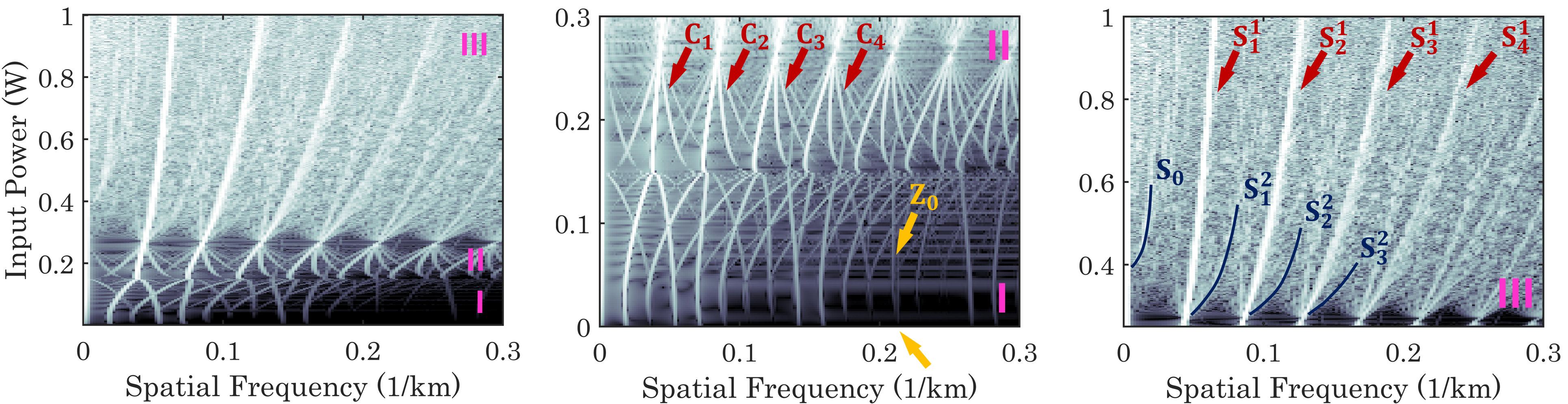}}
\caption{Spatial power spectrum obtained for Soliton Radiation Beat Analysis in a piece of standard single-mode fiber for different input powers of a phase-modulated continuous-wave input. \textit{Left:} full power spectrum; \textit{middle:} detailed view of linear and soliton crystal regimes (I and II); \textit{right:} detailed view of separated soliton regime (III).}
\label{fig:SRBA_opt}
\end{figure*}

Thus, we find: i) there are three input-power dependent regimes that arise out of the initial phase-modulated CW laser input in optical fibers: the quasi-linear temporal Talbot effect for low input powers (I), the nonlinear Talbot effect for intermediate input powers (II), and the regime of separated Talbot solitons for higher input powers (III); ii) in particular, we show that regime II hosts Talbot carpets of soliton crystals rather than rogue waves (cf. \cite{Zhang_2015a}, \cite{Zhang_2015b}, \cite{Nikolic_2019}); iii) the beating of separated Talbot solitons in regime III allows for pulse repetition-rate multiplication; iv) there are two types of solitons: the ones encoded in the whole optical spectrum and, therefore, common for all lines of the frequency comb and the individual solitons carried only by particular comb lines. These results might advance such topics as the increase of the repetition rate of bright and dark pulse trains in fibers and semiconductors (\cite{Cortes_2019}, \cite{Wu_2022}, \cite{Wu_2023}); allow the creation of Talbot soliton fiber lasers (cf. \cite{Zhang_2024}); contribute to the understanding of the dynamics in neuromorphic-photonics schemes that deploy frequency or wavelength multiplexing for information encoding (\cite{Zajnulina_2023}) and deepen the general knowledge about frequency combs in optical fibers.    

\textit{Methods.} To study the temporal Talbot carpets in a passive optical single-mode fiber, we numerically integrate an NLS for the optical field amplitude $A(z,t)$ in the slowly varying envelope approximation in the co-moving frame:
\begin{equation}
   \frac{\partial A}{\partial z} = -i\frac{\beta_2}{2}\frac{\partial^{2}A}{\partial t^{2}} + i\gamma |A|^{2}A
\end{equation}
with $\beta_2 = -23~\text{ps}^{2}/\text{km}$ being the group-velocity dispersion (GVD) parameter and $\gamma = 1.2~\text{W}^{-1}\cdot\text{km}^{-1}$ the nonlinear coefficient at $\lambda_{0} = 1554.6~\text{nm}$. For simplicity, we omit the optical losses, higher-order dispersion, Raman effect, and shock of short pulses usually being present in standard fibers. The initial condition $A(z=0, t)$ replicates the injection of a CW laser with power $P_{0}$ phase-modulated by a modulator with a radio frequency $\Omega = 15.625~\text{GHz}$ and modulation depth $m = 1:$
\begin{equation}
A(z=0, t) = \sqrt{P_{0}}\exp{\left(i\omega_{0} t + im \cos{(2\pi\Omega t)}\right)}.\end{equation}
To highlight the frequency-comb nature of this input, we can rewrite it using the Jacobi-Anger expansion as $A(z=0, t) = \sqrt{P_{0}}\sum_{k = -\infty}^{+\infty} J_{k}(m)\exp{\left(i\omega_{0} t + i k 2\pi\Omega t\right)}$ with $J_{k}(m)$ being a Bessel function of the first kind \cite{Wu_2022}. Thus, this input fulfills the requirement for the temporal Talbot effect \cite{Jannson_1981}. As it enters the fiber, it is subject to the material dispersion that acts as a unitary phase filter inducing the break-up of the field into a train of identical pulses separated by a time period of $\Omega^{-1}$ \cite{Cortes_2019}. In the case of the \textit{linear} temporal Talbot effect, this pulse train will repeat itself with the Talbot period of $L_{T}=\frac{2\pi}{(2\pi\Omega)^{2}|\beta_{2}|} = 28.34~\text{km}.$ 

The SRBA is a numerical technique to retrieve the soliton content from an arbitrary input injected into an optical fiber. It quantitatively analyses spatial frequencies of optical-power oscillations that arise due to one of the following reasons or a combination thereof: higher-order soliton oscillations, beating between several co-existent solitons, and beating between solitons and dispersive waves \cite{Boehm_2006}, \cite{Zajnulina_2015}, \cite{Zajnulina_2017}. Here, we extend this approach by considering the spatial frequencies of the optical-spectrum oscillations over the propagation length. This extension will provide additional information on the soliton types involved.    
%whether the co-existent solitons are separated from each other or rather constitute a soliton crystal, a spatio-temporal soliton compound, as opposed to a Talbot carpet of rogue waves. Also, we will know what parts of the optical spectrum carry the encoded information about the

To perform the SRBA, following steps are to follow: i) numerical integration of the NLS to calculate the optical field and, thus, the optical power over the fiber propagation distance; ii) extraction of the optical power at a time point $t$ for every propagation step, we use $t = 0~\text{ps}$ as the center of the chosen time window; and iii) fast Fourier transform and the calculation of the power spectrum of spatial frequencies. This procedure is repeated for different values of the input power $P_{0}$ and the resulting spectra are stacked into a 2D (spatial frequency - input power) SRBA plot. To interpret this plot, one needs to keep in mind the phase evolution of different kinds of waves in an optical fiber. In general, the phase of a linear (for instance, dispersive) wave evolves mainly depending on GVD and modulation frequency $\Omega:$ $\phi_{lin}(z) \propto\frac{\beta_2\Omega^{2}}{2}z,$ whereas the phase of solitons primarily depends on the input power $P:$ $\phi_{nl}(z) \propto \gamma P z.$ 
%$\phi_{lin}(z) = \frac{\beta_2\Omega^{2}}{2}z,$ whereas the phase of solitons with order $N$ primarily depends on the input power $P:$ $\phi_{nl}(z) = \frac{\left(2N-1\right)\kappa}{2} \gamma P z$ with $\kappa$ being a correction factor. If the whole energy provided to the system is used to form an order-$N$ soliton, $\kappa\approx 1,$ otherwise $\kappa< 1.$
Accordingly, linear waves are recognizable as power-independent \textit{vertical} lines with fixed spatial frequencies, whereas solitons' spatial frequencies constitute input-power dependent parabolic branches in the SRBA plot \cite{Boehm_2006}. This dependence implies that the oscillation period of solitonic waves will nonlinearly decrease with input power, effectively decreasing the initial $L_{T}$ (cf. \cite{Zhang_2015a}).    

Here, we extend the SRBA to consider the spatial oscillations of optical spectra to retrieve additional information about the solitons involved. For this, we perform spatial fast Fourier transform of the optical spectra for chosen values of the input power $P_{0}.$ In such optical frequency - spatial frequency SRBA plots, separated solitons will have well-defined spatial frequencies recognizable as straight \textit{horizontal} lines. These lines correspond to the energy of individual solitons and can be interpreted as energy levels in an atom. For several well-separated solitons, the energy levels will be sharp and quantized in their spatial frequencies. Soliton crystals as soliton compounds will have levels broadened to bands similar to the energy bands of a solid-state crystal. Also, optical frequency - spatial frequency SRBA plots reveal what parts of the optical spectrum carry the information about the solitons and what not. Performing the SRBA for individual frequency comb lines will tell us how lines encode the solitons involved.

\begin{figure}[ht]
\centering
\fbox{\includegraphics[width= 0.4\textwidth]{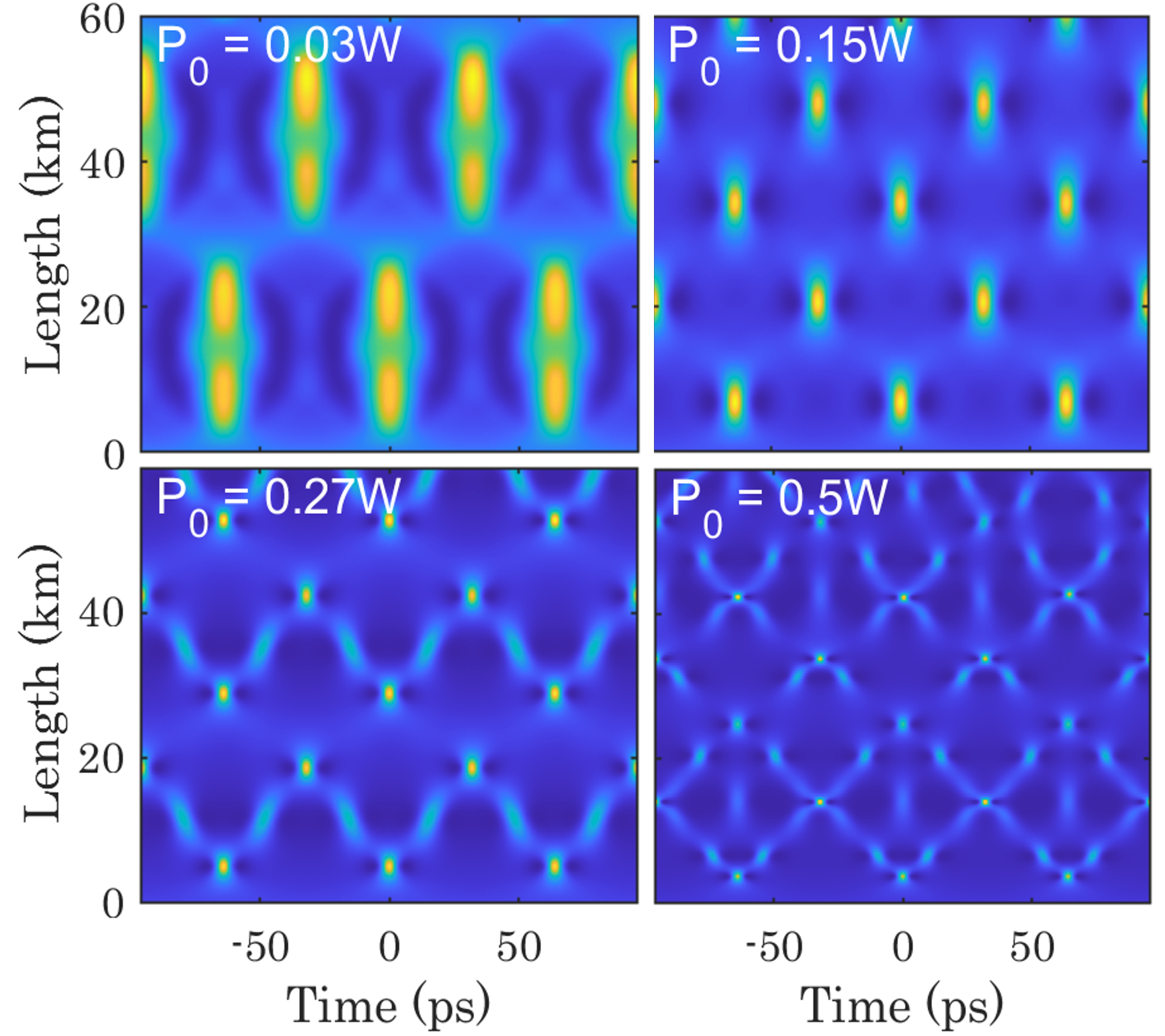}}
\caption{Talbot carpets for different values of the input power $P_{0}$ showing the linear temporal Talbot effect ($P_{0} = 0.03~\text{W}$), the soliton crystals ($P_{0} = 0.15~\text{W}$), separated solitons and the result of their beating ($P_{0} = 0.27~\text{W}$ and $P_{0} = 0.5~\text{W}$).}
\label{fig:opt_power}
\end{figure}

\textit{Results and Discussion.} Fig.~\ref{fig:SRBA_opt} shows an SRBA plot for the fiber parameters mentioned above. We observe three regimes, a quasi-linear one (I) and two nonlinear ones (II and III). The transition between the regimes occurs at well-defined threshold values. Thus, $P_{0} = 0.15~\text{W}$ for the transition from I to II and $P_{0} = 0.27~\text{W}$ for the transition from II to III. The characteristic feature of regime I is the input-power-independent presence of the spatial frequency at $0.213~\text{km}^{-1}.$ With a small difference that arises probably due to numerical imprecision, this value can be associated with the fundamental frequency of the linear Talbot effect: $Z_{0} := \left(2\pi\Omega\right)^{2}|\beta_2| = L_{T}^{-1} = 0.2217~\text{km}^{-1}.$ Also, we see that the Talbot carpet in regime I is created by spatial oscillations with $Z_{n} \approx \frac{Z_{0}}{12}\cdot n,$ $n\in\{1, 2, 3, ...\},$ for $P_{0}\rightarrow 0~\text{W}.$ As the input power increases, these spatial frequencies acquire nonlinear-phase contributions and start growing together under momentum conservation, constituting pitchfork patterns in the SRBA plot (regime I). The corresponding optical field, being subject to modulational instability (MI), gains features that can be associated with a doubly-periodic Akhmediev breather (cf. \cite{Dudley_2009}). In regime II, the latter gives rise to solitonic compounds $C_{l},$ $l\in\{1, 2, 3, ...\}.$ We can associate them with soliton crystals, which we show in more detail in the further course. They are recognizable by regular fans of corresponding sharp, well-defined spatial frequencies. For rogue waves (cf. \cite{Zhang_2015a}, \cite{Zhang_2015b}, \cite{Nikolic_2019}), we would expect washed-out spatial frequencies, which is due to the probabilistic nature of these waves. At the threshold between II and III, the crystals dissolve into separated solitons recognizable by parabolic branches. Each branch $C_{l}$ gives birth to pairs of solitons consisting of a well-pronounced soliton $S_{l}^{1}$ and his less-intensive brother $S_{l}^{2}$ (highlighted in blue for better visibility). At $P_{0}\approx 0.39~\text{W,}$ the energy provided by the input also suffices to create an additional separated soliton $S_{0}$ (cf.~\cite{Zajnulina_2015}). As the appearance and spatial periodicity of these solitons are based on the temporal Talbot effect, we can refer to them as Talbot solitons (cf. \cite{Cohen_2008}).    

Fig.~\ref{fig:opt_power} shows Talbot carpets for different input powers $P_{0}.$ In the quasi-linear regime I, we see patterns of double-hump oscillations that repeat themselves phase-shifted by $\pi$ after $\approx L_{T}$ rather than at $\approx L_{T}/2$ (as it would be the case in the spatial \cite{Wen_2013} or linear temporal Talbot effect \cite{Jannson_1981}). The maximum of the first hump is at $\approx\frac{1}{4}L_{T}$ and of the second at $\approx\frac{3}{4}L_{T}.$ The power minimum between them is at $\approx \frac{1}{2}L_{T}.$ 

\begin{figure}[th]
\centering
\fbox{\includegraphics[width= 0.46\textwidth]{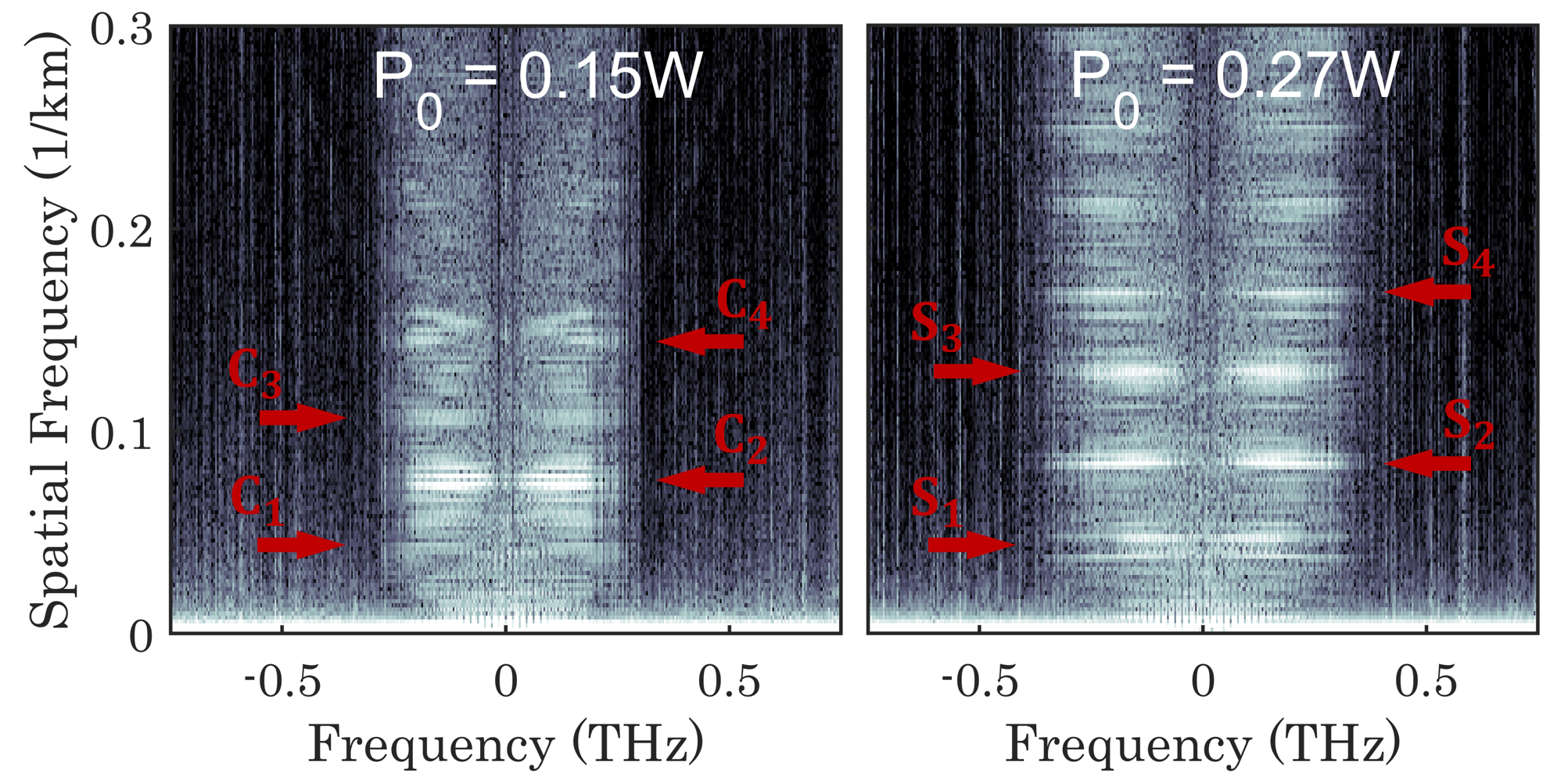}}
\caption{Soliton Radiation Beat Analysis of optical spectra obtained for input powers $P_{0} = 0.15~\text{W}$ and $0.27~\text{W}.$}
\label{fig:SRBA_spec}
\end{figure}

At the I-to-II transition point $P_{0}= 0.15~\text{W},$ we see a carpet that was associated with rogue waves as a result of the nonlinear temporal Talbot effect in Refs.~\cite{Zhang_2015a}, \cite{Zhang_2015b}, \cite{Nikolic_2019}. However, the SRBA reveals that those are soliton crystals $C_{l}$ ( Figs.~\ref{fig:SRBA_opt} and \ref{fig:SRBA_spec}). Also, the power evolution over time and space suggests that we deal here with solitonic waves of a higher order rather than with rogues waves often associated with Peregrine solitons (cf. \cite{Tikan_2017}). In contrast to regime I, the primary pulse-train image repeats itself after a length slightly shorter than $L_{T},$ which is due to the impact of the nonlinearity. The secondary image lies in the middle, replicating the usual spatial and temporal Talbot effect \cite{Wen_2013}, \cite{Jannson_1981}. 

At the II-to-III transition point $P_{0}= 0.27~\text{W},$ we see a nonlinear Talbot carpet of pulses that contain separated Talbot solitons $S_{i}^{1}$ (Fig.~\ref{fig:SRBA_opt}). The carpet consists of primary and secondary images. At $P_{0} = 0.5~\text{W},$ the beating of several separated solitons produces a carpet that resembles fractional (linear) Talbot effect and could be used for pulse repetition-rate multiplication. However, the stability of this carpet over the propagation length is affected due to MI \cite{Nikolic_2019}. 

Fig.~\ref{fig:SRBA_spec} extends the usual SRBA by evaluating spatial frequencies of the optical spectrum at input power values $P_{0}= 0.15~\text{W}$ (I-to-II) and $P_{0}= 0.27~\text{W}$ (II-to-III). Bright horizontal lines indicate the existence of several solitons. In the left figure ($P_{0}= 0.15~\text{W}$), these lines are not sharp but are broadened to bands from which we conclude that the solitons involved are bound to crystals. We can identify bands corresponding to the crystals $C_{1},$ $C_{2},$ $C_{3},$ and $C_{4}$ (cf. Fig.~\ref{fig:SRBA_opt}). The most pronounced is crystal $C_{2}.$ The broadening of the crystal bands is not homogeneous over the optical spectrum but rather happens towards the flanks, i.e. away from $\omega_{0}= 0~\text{THz.}$ In the figure for $P_{0}= 0.27~\text{W},$ we see ca. six different (energy) levels indicating the co-existence of several solitons. Now, the levels are sharper meaning that each level is occupied by a smaller number of separated solitons than in regime II. Thus, $S_{1},$ and $S_{2}$ are visibly split in two, indicating that $C_{1}$ and $C_{2}$ give rise to two solitons each (cf. Fig.~\ref{fig:SRBA_opt}). $S_{3}$ is, however, broad enough to say that it still hosts a crystal rather than several separated solitons. Together with $S_{2},$ $S_{3}$ is the most pronounced structure. Again, the broadening (splitting) of the (energy) levels happens towards the flanks of the optical spectrum. 

\begin{figure}[t]
\centering
\fbox{\includegraphics[width= 0.45\textwidth]{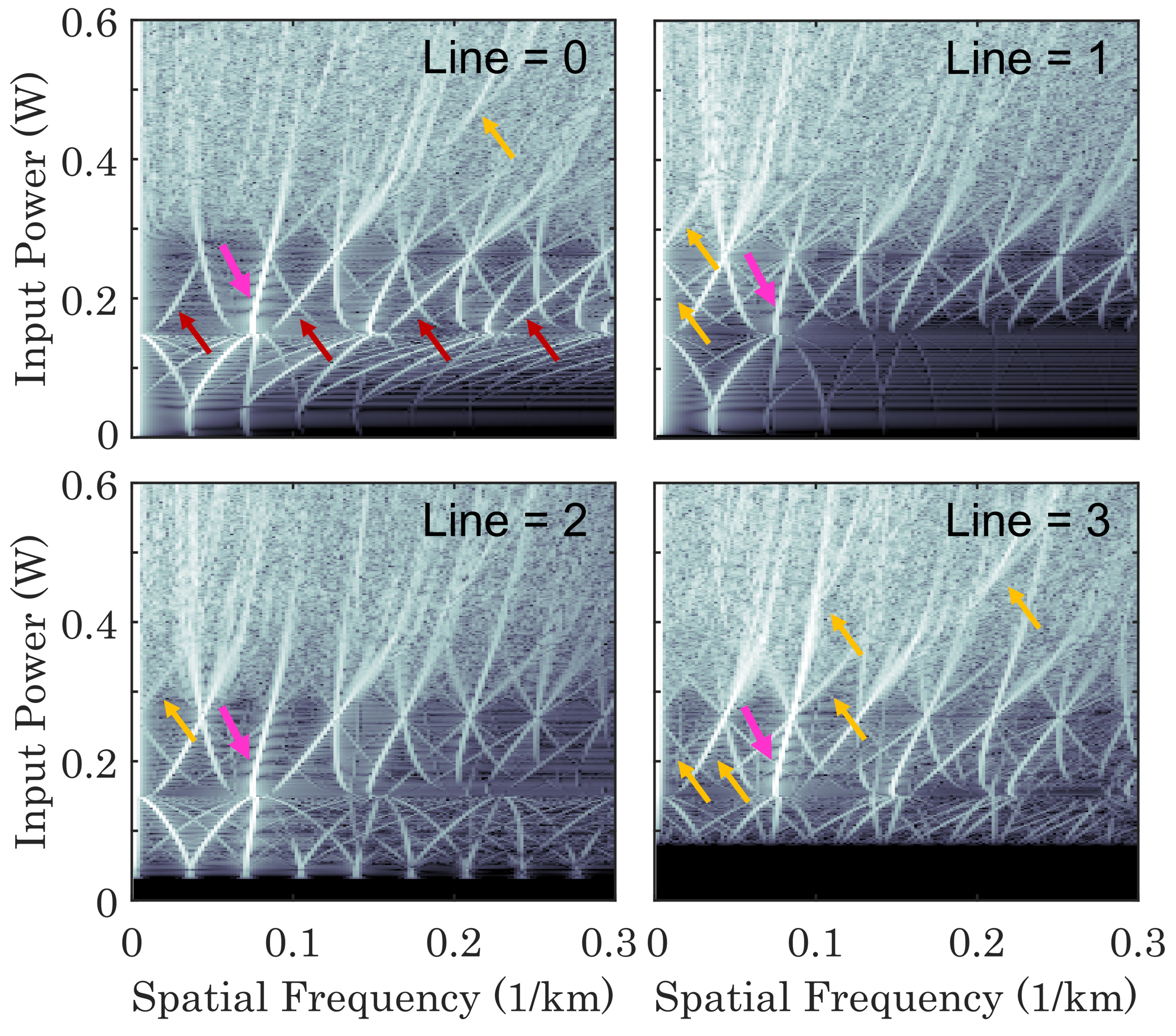}}
\caption{Soliton Radiation Beat Analysis of individual frequency comb lines. Line = 0 denotes the central line at $\omega_{0} = 0~\text{THz},$ the subsequent lines are separated by phase-modulation frequency $\Omega = 15.625~\text{GHz}.$ \textit{Red arrows:} Solitons common for the whole spectrum; \textit{yellow arrows:} line-specific solitons; \textit{magenta arrow:} a quasi-linear wave associated with $C_{2}$ and $S_{2}$; \textit{black horizontal stripes:} input power regions of no spatial oscillations.}
\label{fig:SRBA_lines}
\end{figure}

Fig.~\ref{fig:SRBA_lines} proceeds with SRBA of the optical spectrum. Now, it shows spatial frequencies of the individual frequency-comb lines. Line = j denotes the line at $\omega_{j} := \omega_{0} + j\cdot 2\pi\Omega,$ $j\in\{0, 1, 2, 3\}.$ Here, we see even more co-exiting solitons than recognizable in Fig.~\ref{fig:SRBA_opt}. Some solitons are common for all lines involved (red arrows), and some are encoded only in certain lines (yellow arrows). The further we are away from $\omega_{0}$, the more line-specific solitons appear, explaining the level/band broadening towards the spectrum flanks in Fig.~\ref{fig:SRBA_spec}. In regime II, we now observe many separated solitons encoded in different lines. In general, the lines of a frequency comb are energetically bound. Thus, separated solitons encoded in individual lines are also bound and constitute a soliton crystal (cf. \cite{Zajnulina_2015}, \cite{Zajnulina_2017}). The magenta arrow denotes a quasi-linear wave that can be associated with $C_{2}$ and $S_{2}$ in II and III, respectively. In general, many branches split in regime III indicating radiation of additional solitons from already existing ones. 

\begin{figure}[t]
\centering
\fbox{\includegraphics[width= 0.45\textwidth]{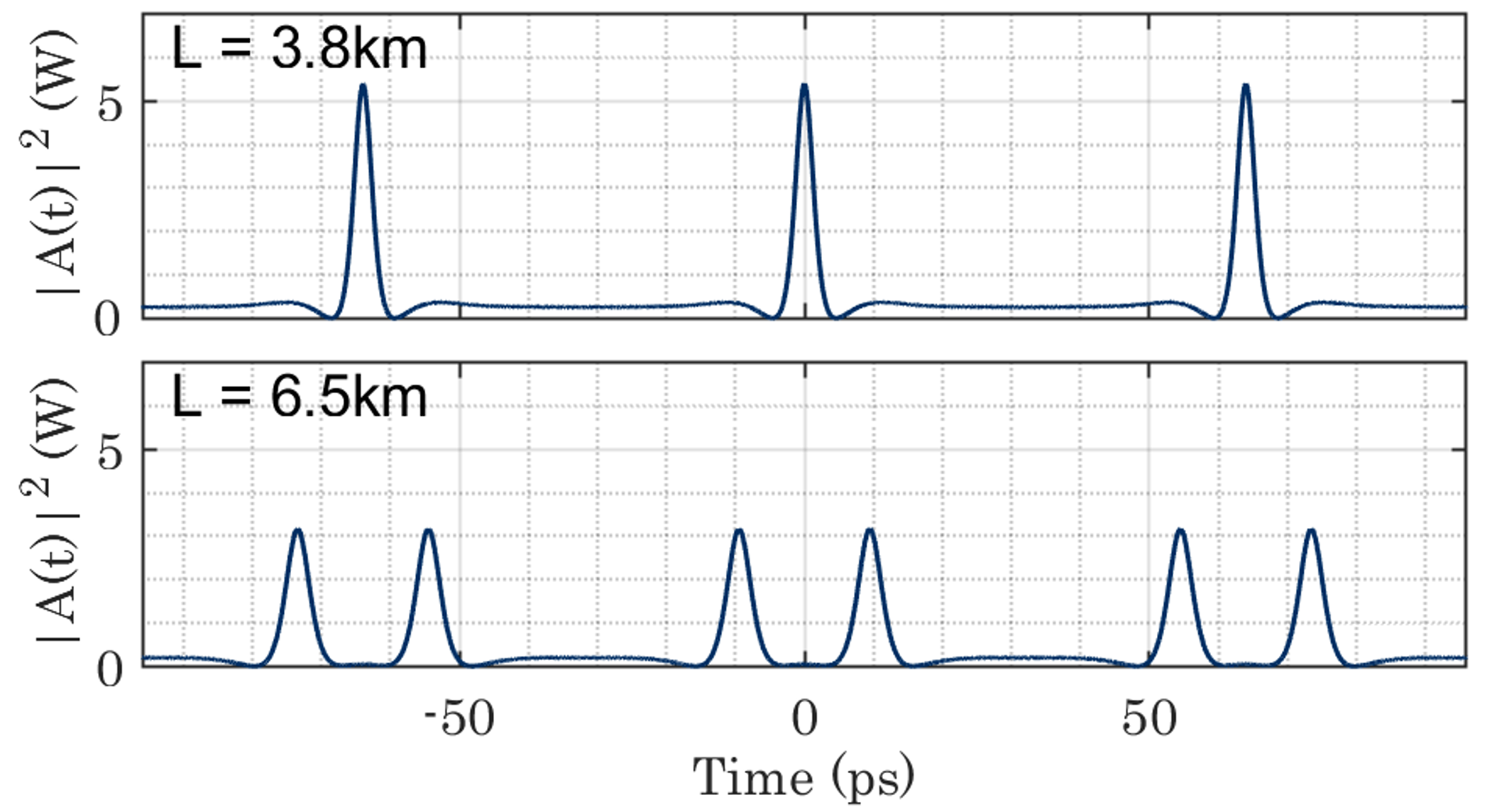}}
\caption{Optical power for $P_{0} = 0.5\text{W}$ (regime III) and fiber lengths $L= 3.8\text{km}$ (primary image as a pulse train with repetition rate $\Omega^{-1})$ and $L=6.5\text{km}$ (pulse multiplication by two).}
\label{fig:PULSE_MULT}
\end{figure}

Finally, Fig.~\ref{fig:PULSE_MULT} shows the primary image of the pulse train that arises for $P_{0}= 0.5~\text{W}$ (regime III). Although the pulses strongly resemble Peregrine solitons, they constitute higher-order Talbot solitons as discussed above (cf. \cite{Tikan_2017}). Their oscillatory behavior can be exploited for pulse repetition-rate multiplication in a highly nonlinear regime. Thus, the pulses are doubled at $L = 6.5~\text{km.}$ However, as we see in Fig.~\ref{fig:opt_power}, the corresponding Talbot carpet is unstable due to the impact of MI (\cite{Nikolic_2019}, \cite{Dudley_2009}). Therefore, only the fiber lengths within the first reoccurrence of the primary image should be considered for pulse repetition-rate increase.

\textit{Conclusion.} Using the numerical technique of the Soliton Radiation Beat Analysis and a phase-modulated CW field as an initial condition for the Nonlinear Schr{\"o}dinger equation, we studied the quasi-linear and nonlinear temporal Talbot effect in optical fibers. We found the input-power-dependent threshold between the quasi-linear Talbot effect and the nonlinear Talbot effect of soliton crystals, as well as between the crystals and Talbot solitons. We showed that the beating of Talbot solitons allows for pulse repetition-rate multiplication. Also, we observed two types of solitons involved: the ones encoded in the whole optical spectrum and the ones carried by particular comb lines. These studies contribute to the understanding of the evolution of solitonic pulse trains and the corresponding frequency combs in optical fibers. However, deeper studies are needed to precisely understand the formation of soliton crystals under the impact of the temporal Talbot effect. 

%Further studies are needed to better understand soliton and soliton-crystal evolution in the context of temporal Talbot effect. In particular, such effects as higher-order dispersion, shock, optical losses, and Raman effect should be taken into account to replicate optical-field evolution in a real fiber.  

\begin{backmatter}
\bmsection{Funding} We acknowledge the funding from the project Win4Space being a part of the Win4ReLaunch initiative supported by SPW Economie Emploi Recherche of the Walloon Region Belgium (grant agreement number 2210181).

\bmsection{Disclosures} We declare no conflict of interest.

\bmsection{Data availability} Data underlying the results presented in this paper are not publicly available at this time but may be obtained from us upon reasonable request.

\end{backmatter}

% Bibliography
\bibliography{sample}

% Full bibliography added automatically for Optics Letters submissions; the following line will simply be ignored if submitting to other journals.
% Note that this extra page will not count against page length
\bibliographyfullrefs{sample}

%Manual citation list
%\begin{thebibliography}{1}
%\bibitem{Zhang:14}
%Y.~Zhang, S.~Qiao, L.~Sun, Q.~W. Shi, W.~Huang, %L.~Li, and Z.~Yang,
 % \enquote{Photoinduced active terahertz metamaterials with nanostructured
  %vanadium dioxide film deposited by sol-gel method,} Opt. Express \textbf{22},
  %11070--11078 (2014).
%\end{thebibliography}

% Please include bios and photos of all authors for aop articles
\ifthenelse{\equal{\journalref}{aop}}{%
\section*{Author Biographies}
\begingroup
\setlength\intextsep{0pt}
\begin{minipage}[t][6.3cm][t]{1.0\textwidth} % Adjust height [6.3cm] as required for separation of bio photos.
  \begin{wrapfigure}{L}{0.25\textwidth}
    \includegraphics[width=0.25\textwidth]{john_smith.eps}
  \end{wrapfigure}
  \noindent
  {\bfseries John Smith} received his BSc (Mathematics) in 2000 from The University of Maryland. His research interests include lasers and optics.
\end{minipage}
\begin{minipage}{1.0\textwidth}
  \begin{wrapfigure}{L}{0.25\textwidth}
    \includegraphics[width=0.25\textwidth]{alice_smith.eps}
  \end{wrapfigure}
  \noindent
  {\bfseries Alice Smith} also received her BSc (Mathematics) in 2000 from The University of Maryland. Her research interests also include lasers and optics.
\end{minipage}
\endgroup
}{}

\end{document}